\documentclass[showpacs,aps,prb,twocolumn,floatfix]{revtex4}
\usepackage{graphicx}
\usepackage{bm}
\usepackage{amsmath}

\newcommand{\s}{\sum\limits}
\newcommand{\p}{\prod\limits}
\newcommand{\pa}{\partial}
\newcommand{\il}{\int\limits}
\newcommand{\be}{\begin{equation}}
\newcommand{\e}{\end{equation}}
\newcommand{\beml}{\begin{subequations}}
\newcommand{\eml}{\end{subequations}}
\newcommand{\beq}{\begin{eqnarray}}
\newcommand{\eq}{\end{eqnarray}}
\newcommand{\ba}{\begin{array}}
\newcommand{\ea}{\end{array}}
\newcommand{\lt}{\left}
\newcommand{\rt}{\right}
\newcommand{\n}{\nonumber}
\newcommand{\la}{\langle}
\newcommand{\ra}{\rangle}
\newcommand{\tr}{{\rm Tr}\,}

\newcommand{\diag}{\,{\rm diag}\,}
\newcommand{\im}{\,{\rm Im}\,}
\newcommand{\re}{\,{\rm Re}\,}
\newcommand{\ep}{\varepsilon}

\begin{document}

\date{August 7, 2002}

\title{Random matrix theory of proximity effect in disordered wires}

\author{M. Titov}
\author{H. Schomerus}

\affiliation{Max-Planck-Institut f\"ur Physik komplexer Systeme,
N\"othnitzer Str. 38, 01187 Dresden, Germany
}

\begin{abstract}
We study analytically the local density of states (LDOS) 
in a disordered normal-metal wire (N) at ballistic distance to a  
superconductor (S). Our calculation is based on a 
scattering-matrix approach, which concerns for wave-function
localisation in the normal metal,
and extends beyond the conventional semiclassical theory based on Usadel 
and Eilenberger equations.  
We also analyse how a finite transparency of the NS interface
modifies the spectral proximity effect and demonstrate
that our results agree in the dirty diffusive limit
with those obtained from the Usadel equation.
\end{abstract}
\pacs{
74.50.+r, %Proximity effects, weak links, tunnelling phenomena, 
%and Josephson effects  
74.80.Fp, %Point contacts; SN and SNS junctions
73.20.Fz %Weak or Anderson localisation
%73.20.At %Surface states, band structure, electron density of states
%73.63.Rt, %Nanoscale contacts
72.15.Rn  %Localisation effects (Anderson or weak localisation)
}

\maketitle
\section{Introduction}

It is widely acknowledged that a piece of a normal metal 
that is in good contact with a superconductor acquires some
superconducting properties.
This phenomenon, named the proximity effect, 
has already been studied by Cooper\cite{Cooper}
in the early sixties. Since then many 
theoretical and experimental investigations have been
carried out.\cite{DeuGennes}
Much owed to the recent progress in the fabrication technology 
of nanostructures there is a revived interest 
to the proximity effect in the last decade.\cite{KSS} 
One remarkable evidence of this effect is the
formation of a spectral gap in the normal metal, 
which strongly affects the 
low-temperature transport properties of the normal-metal
superconductor (NS) junctions. 
The key mechanism responsible for the appearance of the gap
is the Andreev reflection at the NS boundary, which converts the
dissipative electrical current into dissipationless 
supercurrent.\cite{A}
Similar mechanisms act in superconductor ferromagnet junctions
which have become an object of intense study 
recently.\cite{BVE_BB_HE_FL,SHLH}

An effective experimental technique which allows for
spatially resolved measurements of the electronic density
in the nanostructures is the scanning tunnelling microscopy. 
It provides both a unique sub-meV energy sensitivity and 
an atomic spatial resolution. 
Several recent measurements of the local 
electronic density of states (LDOS) in 
the NS junctions\cite{MCP,VCL,GPBED,THL_TTHMG,LMRPM}
turned out to be in very good agreement with the predictions of
quasiclassical theory\cite{KL,GK88,GK96,BBS,PBB,FF} 
of ``non-equilibrium'' superconductivity,
based on the Usadel equation for the diffusive transport\cite{U}
and the Eilenberger equation for the ballistic transport.\cite{E} 

The interplay of ballistic and diffusive transport becomes 
important when one studies local properties at short distance to
an NS interface in a disordered system. Quasiparticles are then
transferred to the interface by ballistic transport, while they explore 
the rest of the system diffusively. This situation is not covered
by conventional quasiclassical theory. Quasiclassics also cannot
account for the non-perturbative effects of wave-function localisation,
which only can be included by a fully phase-coherent approach.

In this paper we present a theory that goes beyond the quasiclassical 
description and apply it to calculate the local density of states
in an NS wire geometry near the interface, at zero temperature and
vanishing magnetic field.

In our model the normal metal is shaped in the form of the
long disordered quantum wire, 
which supports $N$ propagating modes at the 
Fermi level $E_F$. The elastic scattering mean free path $\ell$
in the wire is assumed to be much larger than the 
Fermi wave length $\lambda_F$, which corresponds to the weak 
disorder.
The superconductor is assumed to be clean and characterised by 
the bulk value $\Delta$ of the amplitude of the pair potential.
The superconductor order parameter is assumed to be constant $\Delta$
in the superconductor and zero in the normal metal.
This approximation is 
referred to in the literature as a ``rigid boundary condition''.\cite{L} 

\begin{figure}
\includegraphics[width=6cm]{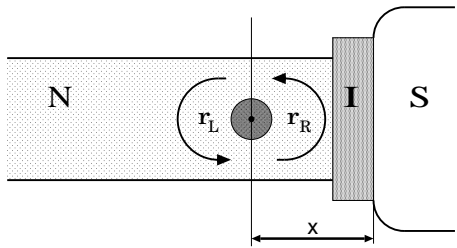}
\caption{
The geometry of an NS junction consisting of a 
long normal-metal disordered wire N,
a clean superconductor S
and a dielectric tunnel barrier I in between. 
The mean local density of states (LDOS) is calculated at the 
distance $x$ from the NS interface, with 
$\lambda_F \ll x \ll \ell$.
The matrix $r_L$ relates the plane-wave components 
in the process of reflection from the 
normal-metal disordered wire. 
The matrix $r_R$ describes the reflection from the 
tunnel barrier -- superconductor part of the junction.
The mean LDOS is found by averaging over the 
disorder-induced fluctuations of the matrix $r_L$.
}
\label{fig:setup}
\end{figure}

We calculate the mean LDOS, or, more precisely, its envelope,
at a distance $x$ on the normal-metal side of the NS interface 
as shown schematically in Fig.~\ref{fig:setup}.
The envelope is obtained by averaging the LDOS 
over distances of the order of the Fermi wave length $\lambda_F$. 
The spatial averaging smears out the 
Friedel type oscillations and makes the LDOS
independent on the position across the wire.

We study in detail the case that
the distance $x$ is small compared to the scattering mean free path
$\ell$, so that $\lambda_F \ll x \ll \ell$, 
while the ratio between the superconductor coherence length
$\xi=\hbar v_F/\Delta$ and $\ell$ remains arbitrary. 
The resulting mean LDOS found by averaging over disorder 
does not depend on $x$ and is a smooth function of energy 
everywhere except at $\ep=\Delta$ (the energy $\ep$ is
measured from the Fermi surface). 

Our calculation is organised as follows. In Sec.\ \ref{sc:Green}
we derive a general relation between the one-point Green function 
in a quantum wire and the reflection matrices $r_L$, $r_R$.
These matrices relate the plane-wave components 
of the quasiparticle wave function 
in the process of reflection from the parts of the wire to the 
left and to the right part of $x$.

We apply this result in Sec.\ \ref{sc:LDOS} in order to calculate the mean
LDOS in the neighbourhood of an ideally transmitting NS interface.
The matrices $r_L$, $r_R$ of the size $2N\times 2N$
describe the reflection of the electron- and hole-like quasiparticles. 
The left reflection matrix $r_L$ is diagonal in the electron-hole
representation and depends on the disorder in the normal metal.
The right reflection matrix $r_R$ is off-diagonal (in absence of the
tunnel barrier) and is fixed within the model considered.

%%%%%%%%%%%%%%%%%%%%%%%%%%%%%%%
%%%%%%%%%%%%%%%%%%%%%%%%%%%%%%%%%

\begin{figure}
\includegraphics[width=7.5cm]{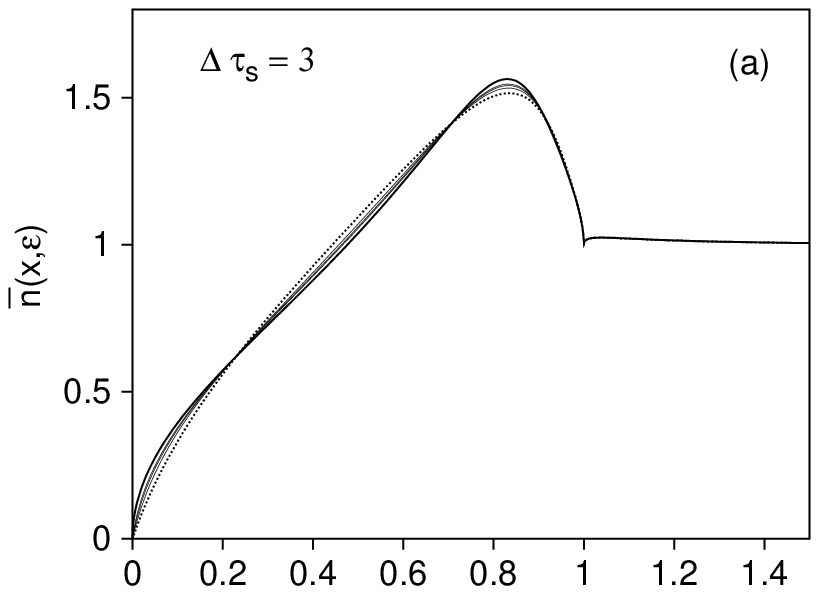}
\includegraphics[width=7.5cm]{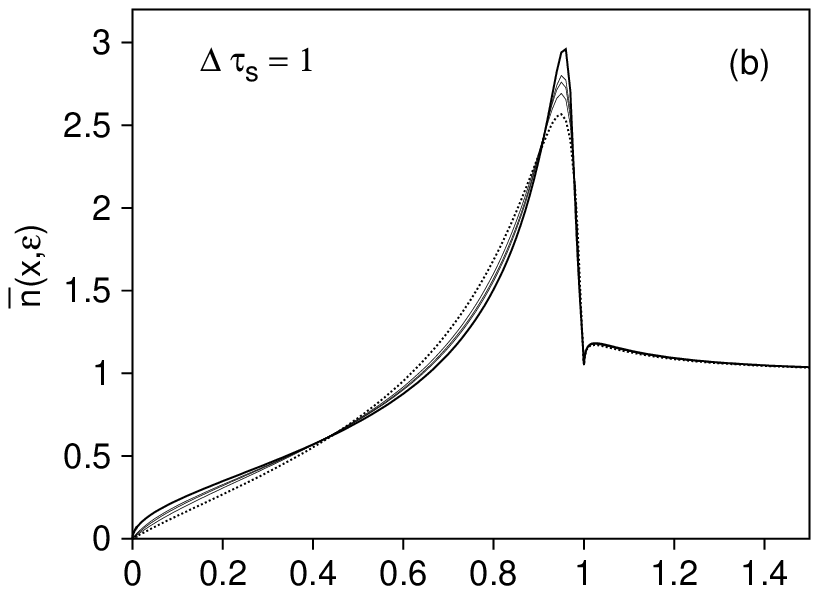}
\includegraphics[width=7.5cm]{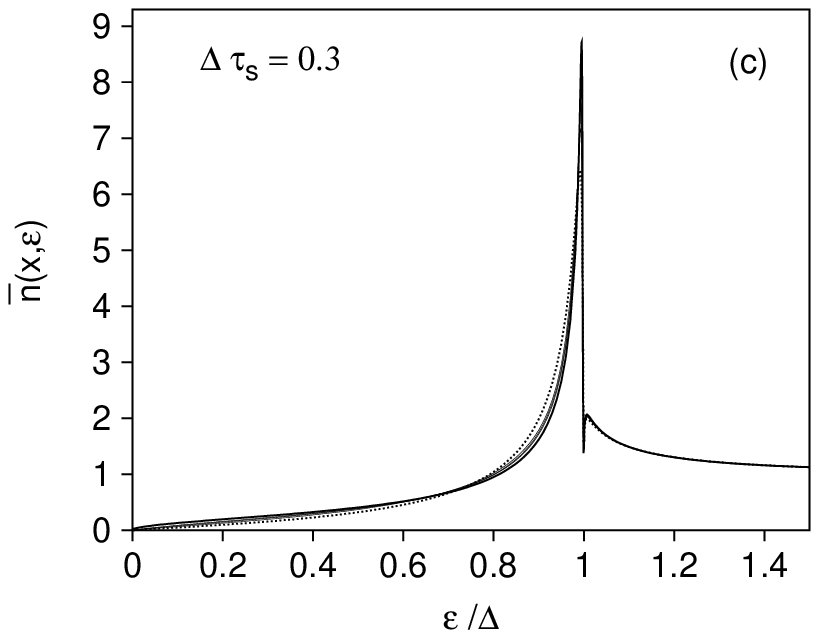}
\caption{
The mean LDOS (\ref{result0}) 
in an N-channel normal-metal wire
near an ideally transmitting NS
interface. The curves are calculated
from Eqs.\ (\ref{orthogonal},\ref{result0}). 
The thick solid (dotted) line corresponds to the 
limiting case of the multi(single)-channel 
wire. The thin lines are for the finite number 
of channels $N=2$, $3$, and $4$. The figures
correspond to the clean regime $\Delta\tau_s=3$ (top),
the intermediate regime $\Delta\tau_s=1$ (middle), and
the dirty regime $\Delta\tau_s=0.3$ (bottom).
}
\label{fig:figT1}
\end{figure}

In the region $\ep < \Delta$ we obtain the 
disorder-averaged LDOS 
\be
\label{first}
\bar{n}(x,\ep)=\pi \rho_{\ep}(\phi_A), \qquad \phi_A=\arccos{\ep/\Delta}, 
\e
where the function $\rho_\ep(\phi)$ is 
the probability density of the eigenphase of the matrix correlator
$r_0(\ep)r_0(-\ep)^\dagger$.
The reflection matrix $r_0(\ep)$ 
relates the plane-wave amplitudes of the electron wave function
in the process of reflection from the semi-infinite normal-metal wire.
The probability density $\rho_\ep(\phi)$ has been studied in
Ref.\ \onlinecite{TB}. Apart from energy and phase it depends on
the number of channels $N$ and the mean scattering time $\tau_s=\ell/v_F$. 
According to Ref.\ \onlinecite{TB} one can distinguish
localised, diffusive and ballistic regimes in the form of the function
$\rho_\ep(\phi)$ depending on the value of $\ep$. 
We observe the effect of Anderson localisation 
in the linear increase of the LDOS for energies 
smaller than the Thouless energy $\ep_c=\hbar/N^2\tau_s$.
We also find that the curves calculated 
for different number of channels in the wire 
are lying close to each other at any ratio $\ell/\xi$
(see Fig.~\ref{fig:figT1}). This suggests that the weak-localisation 
correction to the LDOS 
is small in the case of the ideally transmitting NS interface.

In Sec.\ \ref{sc:Effect} we generalise the model to include
a tunnel barrier at the interface,
parametrised by a tunnel probability per mode $\Gamma$. 
We calculate analytically the LDOS near the interface
in the extreme cases of a localised wire, $N=1$, and 
a diffusive wire, $N\gg 1$.

\begin{figure}
\includegraphics[width=7.5cm]{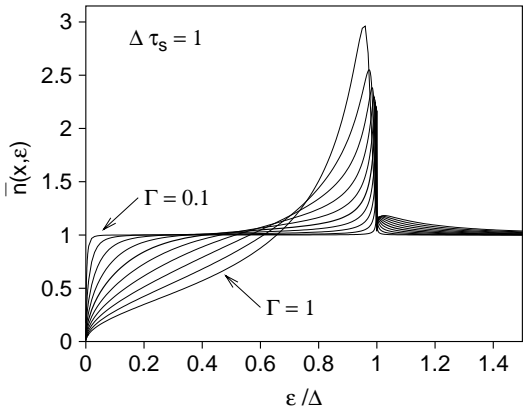}
\caption{
The mean LDOS in a diffusive normal-metal wire
in the vicinity of an NS interface of finite
transparency. The curves are calculated 
from Eqs.\ (\ref{exciting},\ref{InftymodeLDOS})
for $\Delta \tau_s=1$ and 
the effective tunnelling probability $\Gamma$
varying from $0.1$ to $1$ in steps of $0.1$.
The thick line corresponds to $\Gamma=0.5$.
}
\label{fig:figTrange}
\end{figure}

The effect of the tunnel barrier 
consists of a reduction of the 
pseudogap in the normal metal. This effect
is most pronounced in the dirty regime 
$\ell \lesssim \xi$, or $\Delta \tau_s/\hbar \lesssim 1$.
The results of our calculation for the diffusive wire
in the intermediate regime $\ell=\xi$
are summarised in Fig.~\ref{fig:figTrange}
for different values of $\Gamma$.  
We observe that the
LDOS increases monotonously to its bulk constant value 
around the energy $\hbar \Gamma^2/\tau_s$ 
and reveals a high and narrow peak close to $\ep=\Delta$.

The monotonous reduction of the pseudogap 
is attributed to the quasiparticles which experience 
normal reflection at the tunnel barrier and therefore 
do not see the NS boundary. 
The formation of the peak is due to the 
quasiparticles reflected from the superconductor.

When the distance $x$ increases beyond 
the mean free path $\ell$ a 
competing effect takes place. That is the suppression of the pseudogap 
due to the back-scattering on the weak disorder potential 
in the normal-metal segment of length $x$
in front of the interface. 
The estimated size of the pseudogap due to this effect
is $\hbar D/x^2$, where $D$ is the diffusion constant 
in the normal metal. In this case the LDOS considerably
overshoots its bulk value around $\ep =\hbar D/x^2$,
which is in contrast to the monotonous increase
due to the tunnelling into the superconductor.  
We therefore anticipate that the effect of the tunnel barrier
still can be seen in the shape of the LDOS provided
$\hbar D/x^2 \gg \hbar \Gamma^2/\tau_s$, or equivalently
$x \ll \ell/\Gamma$. Namely, at distances smaller than 
$\ell/\Gamma$ the LDOS may acquire the step-like feature
at the value $\ep =\hbar \Gamma^2/\tau_s$, which is fixed  
by the NS interface transparency rather than 
by the distance to the interface. 

A qualitatively similar phenomenon 
has been indeed observed in experiments by 
Levi {\it et. al.}\cite{LMRPM} in the Cu barrier pin wires
near a N(Cu)-S(NbTi) boundary. 

On the contrary, at large 
distances $x\gg \ell/\Gamma$ the barrier is not effective in the sense
that its presence cannot be distinguished in the 
energy dependence of the LDOS. 
This is consistent with a general 
semiclassical criterion\cite{SIL} which states that the barrier 
is not effective for a given observable
if the most of the relevant trajectories 
hit the NS interface more than $\Gamma^{-1}$ times before 
the electron-hole coherence is lost. In the case of the LDOS
this criterion is fulfilled for $x\gg \ell/\Gamma$.

The tunnel barrier acts differently for the single-channel wire. 
In the dirty regime $\ell \lesssim \xi$
the size of the pseudogap $\hbar \Gamma/\tau_s$ 
scales linearly with $\Gamma$ 
due to the Anderson localisation.
This results in a different shape of the LDOS
compared to the diffusive case ($N\gg 1$). The difference
becomes more and more
pronounced with decreasing ratio $\ell/\xi$ or tunnelling 
probability $\Gamma$. 

In Sec.\ \ref{sc:Usadel} we compare the LDOS for the
diffusive case $(N\gg 1)$ found from our theory to
the LDOS calculated from the Usadel equation.\cite{GK96} 

\section{Green function in a wire geometry} \label{sc:Green}

In our model of the NS junction the normal metal
is shaped in the form of a semi-infinite
quasi-one dimensional disordered wire.
The properties of such a system is well 
understood in the framework of the scattering theory\cite{CWJ}
provided the weak disorder limit $\lambda_F \ll \ell$.
The detailed statistical description of the 
disorder scattering is based on the Dorokhov-Mello-Pereyra-Kumar 
(DMPK) equation.\cite{D,MPK} This is a scaling equation 
for the probability distribution of
the scattering matrix of a segment of the wire. 
Below we derive a general relation between the
one-point Green function and the reflection matrices
$r_L$, $r_R$ for two parts of the wire. The single-channel 
counterpart of this
relation has been used recently to reconsider the problem of
LDOS fluctuations in 1D normal-metal wires.\cite{STBB} 

The disordered wire has the Hamiltonian 
$H=H_0+V(\vec{r})$, where $V(\vec{r})$ is a disordered
potential. We parametrise $\vec{r}=(x,\vec{\rho})$, where
$x$ is the coordinate along the wire and $\vec{\rho}$ is the vector
in the transversal direction. We first discuss 
the case of `spinless' electrons, assuming 
$H_0=-\frac{1}{2 m_e} \nabla^2$, $\hbar=1$, and include
hole-like quasiparticles in Sections III and IV.
 
In the absence of $V$ the quantisation in the transversal 
direction gives rise to a set of $N$ propagating modes
characterised by the transverse momentum ${\vec{q}}_n$.
The total energy $E=(1/2m_e)(|{\vec{q}}_n|^2+k_n^2)$, 
where the $x$-momentum $k_n$ is conserved.
The retarded Green function
$G^R(E)=(E+i\eta -H)^{-1}$ is written
in the channel representation as
\be
\label{green}
G^R_{nm}(x,x^\prime)=\int\!\!\!\int_A d\vec{\rho}\, 
d\vec{\rho}^{\;\prime}
\la n |\vec{\rho} \ra
\la \vec{\rho}^\prime| m \ra 
\la\vec{r}|G^R|\vec{r}^{\;\prime}\ra,
\e
where the integration is carried out over 
a cross-sectional area $A$. Hence the LDOS 
\be
\label{def_nu_mic}
n(\vec{r},\ep)=-\frac{1}{\pi} \im \s_{n, m} 
\la m |\vec{\rho} \ra
\la \vec{\rho}| n \ra 
G_{nm}^R(x,x),
\e
where $\ep$ is the energy measured from the Fermi surface.
For a two-dimensional wire of the width $d$ we
have $\la \rho | n\ra=(2/d)^{1/2}\sin{(\pi n \rho/ d)}$. 
In what follows we shall omit the index $R$, assuming everywhere
the retarded Green function.

Let us formally cut the wire in the point $x$ into two pieces
and treat the left and the right part separately.
We decompose the potential $V=V_R+V_L$, where
$V_{R,L}$ is the disorder potential in the right 
and the left part of the wire, respectively.
We also introduce the left and the right Green function
as $G_{R,L}=(E+i\eta-H_0-V_{L,R})^{-1}$. 
According to Fisher and Lee, Ref.\ \onlinecite{FL}, we have
\be
\label{Fisher_Lee}
G_{L,R;nm}(x,x)=
\frac{1}{i\sqrt{v_n v_m}}(\delta_{nm}+r_{L,R;nm}(x)),
\e
where $v_n=k_n/m_e$ is the channel velocity and
$r_{L,R}$ are the reflection matrices 
from the left and 
the right part of the wire, respectively.

The Green functions obey Dyson equations
which can be written in the matrix form as
\beml
\label{Dyson_eqs}
\beq
\hspace*{-1cm}
&&\hat{G}(x,x)=\hat{G}_{0}(x,x)+\!\!\int_{-\infty}^\infty 
\!\!\!dy\; \hat{G}_{0}(x,y)\hat{V}(y) \hat{G}(y,x),\\
\hspace*{-1cm}
&&\hat{G}(x,x)=\hat{G}_{R}(x,x)+\!\!\int_{-\infty}^x 
\!\!\!dy\; \hat{G}_{R}(x,y)\hat{V}_{L}(y)  \hat{G}(y,x),\\
\hspace*{-1cm}
&&\hat{G}(x,x)=\hat{G}_{L}(x,x)+\!\!\int_x^\infty 
\!\!\!dy\; \hat{G}_{L}(x,y)\hat{V}_{R}(y)\hat{G}(y,x),
\eq
\eml
where the elements of the matrix $\hat{V}$
are given by
\be
V_{nm}(x)=\int_A d\vec{\rho} \; 
\la n| \vec{\rho}\ra \la \vec{\rho}| m\ra\,V(\vec{r}),
\e
and the ballistic Green function 
(in absence of the potential) reads
\be
\label{G0}
G_{0,nm}(x,x^\prime)=\frac{\delta_{nm}}{i v_n} e^{i k_n |x-x^\prime|}.
\e
We also take advantage of the following relations\cite{GCB}
\beml
\beq
\hspace*{-1cm}
G_{R,nl}(x,y)&=&e^{-ik_l(x-y)}G_{R,nl}(x,x), \quad \mbox{for }y<x,\\
\hspace*{-1cm}
G_{L,nl}(x,y)&=&e^{-ik_l(x-y)}G_{L,nl}(x,x), \quad \mbox{for }y>x,
\eq
\eml
in the disorder-free regions in order to 
eliminate the integral terms in Eq.\ (\ref{Dyson_eqs}).
As a result we obtain the matrix equality
\be
\frac{1}{\hat{G}(x,x)}+\frac{1}{\hat{G}_0(x,x)}=
\frac{1}{\hat{G}_R(x,x)}+\frac{1}{\hat{G}_L(x,x)}.
\e
Using Eqs.\ (\ref{Fisher_Lee}) and (\ref{G0}) we finally get
\be
\label{gen_result}
\hat{G}(x,x)
=\frac{1}{\sqrt{i\hat{\rm{v}}}}
(1+\hat{r}_R)\frac{1}{1-\hat{r}_L \hat{r}_R}
(1+\hat{r}_L)
\frac{1}{\sqrt{i\hat{\rm{v}}}},
\e
where $\hat{\rm{v}}$ is the diagonal matrix
of channel velocities $v_n$. 
Together with Eq.\ (\ref{def_nu_mic}) this equation
defines the LDOS via the reflection matrices.

In the case of uncorrelated disorder
the reflection matrices $r_L$ and $r_R$ are 
statistically independent, which makes 
Eq.\ (\ref{gen_result}) useful for practical calculations.

In general the LDOS oscillates on the
scale of $\lambda_F$ (due to the prevailing contribution
of one particular quantum state). 
These Friedel-type oscillations can play a crucial role 
especially in one dimension. 
In what follows we are concerned with the
smoothed version of the LDOS that does not change on the scale of
the Fermi wave length and, therefore, also not in the 
transversal direction. For this purpose
we introduce the spatially averaged LDOS
\be
n(x,\ep)=\delta V^{-1}\int_{\delta V}n(\vec{r},\ep)\,d\vec{r},
\e
where the integration is carried out over a volume 
$\delta V$ around the point $(x,\vec{\rho})$. 
The linear size of the volume $\delta V$ is assumed to be
much larger than the Fermi wave length and much smaller
than the mean free path $\ell$.
For $|x-x^\prime| \ll \ell$
the reflection matrices defined at the cross section
$x^\prime$ are related to those defined at $x$ by 
\beml
\label{shift}
\beq
r_{L}(x^\prime)&=&e^{-i\hat{k}(x-x^\prime)}r_{L}(x)e^{-i\hat{k}(x-x^\prime)},\\
r_{R}(x^\prime)&=&e^{i\hat{k}(x-x^\prime)}r_{R}(x)e^{i\hat{k}(x-x^\prime)},
\eq
\eml
with $\hat{k}=m_e \hat{{\rm v}}$. Expanding the right-hand side
of Eq.\ (\ref{gen_result}) in a geometric series 
in $r_L$, $r_R$ we notice that only the terms with 
equal numbers of $r_L$ and $r_R$ matrices do not oscillate
on the scale of the Fermi wave length
and have to be kept. Additionally the averaging in 
the transversal direction mixes up the different modes
so that 
$\la m| \vec{\rho} \ra \la \vec{\rho} | n \ra \propto \delta_{mn}$
in Eq.\ (\ref{def_nu_mic}). As the result we obtain
\be
\label{meso}
n(x,\ep)=\frac{n_0}{N} \re \tr 
\frac{1+r_R r_L}{1-r_R r_L},
\e
where $n_0$ is the bulk value of the LDOS
in the normal metal, which is set to unity in the rest of the paper.
In what follows we apply Eq.\ (\ref{meso})
to calculate the LDOS in the normal-metal wire
in the immediate vicinity of an NS interface. 

%%%%%%%%%%%%%%%%%%%%%%%%%%%%%%%
%%%%%%%%%%%%%%%%%%%%%%%%%%%%%%%%%

\section{LDOS near the ideal NS interface} \label{sc:LDOS}

The relation (\ref{meso}) applies straightforwardly
to the model of the NS junction discussed 
in the Introduction. The only modification is 
the doubling of size of the reflection matrices due to 
particle-hole conversion. We still denote 
the number of electron channels in the wire by $N$,
so that the size of the 
particle-hole reflection matrix is now $2N$.
Equation (\ref{meso}) can be written in the form
\be
\label{meso2}
n(x,\ep)=1+\frac{2}{2N} \re \tr \s_{n=1}^\infty
(r_L r_R)^n,
\e
where $r_L$ is the electron-hole reflection matrix 
for the long normal-metal wire, 
while $r_R$ is that for the ideal NS interface.
These reflection matrices are conveniently parametrised by
\be
\label{par1}
r_L=\lt(
\ba{cc}
r_0(\ep)&0\\ 0& r_0(-\ep)^\ast 
\ea
\rt), \quad
r_R=e^{-i\phi_A}\lt(
\ba{cc}
0&1\\ 1&0
\ea
\rt),
\e
where $\phi_A=\arccos\ep/\Delta$ is the Andreev phase
and $r_0(\ep)$ [$r_0(\ep)^\ast$]
is $N\times N$ reflection matrix of the electron-like
[hole-like] quasiparticles for 
the normal-metal wire.
The matrix product $r_L r_R$ is block off-diagonal,
hence only the even powers $n$ contribute to the trace in 
Eq.\ (\ref{meso2}). 
From Eqs.\ (\ref{meso2},\ref{par1}) we obtain 
\be
\label{easy}
n(x,\ep)=1\!+\!\frac{2}{N}\re \tr \s_{n=1}^\infty
\lt[r_0(\ep)r_0(-\ep)^\ast\rt]^n e^{-2 i n \phi_A}.
\e 
The right-hand side of Eq.\ (\ref{easy}) 
is completely determined by the eigenvalues of the  
correlator $r_0(\ep)r_0(-\ep)^\ast$, 
which is a unitary matrix. 
Its eigenvalues are conveniently parametrised
by $\exp(2 i \phi_j)$, $j=1, 2, \dots, N$,
where the phases $\phi_j$ are restricted to the interval
$(0,\pi)$. The joint probability density 
$P_\ep(\phi_1, \phi_2, \dots, \phi_N)$ is 
a symmetric function with respect to any permutation of its arguments
because of the statistical equivalence of the channels.
This function has been studied in detail in Ref.\ \onlinecite{TB}. 
Our calculation is restricted to the mean LDOS
$\bar{n}(x,\ep)
\equiv \la n(x,\ep) \ra$, where the angular brackets correspond 
to the average over the disorder potential in the wire.
In order to perform the average in Eq.\ (\ref{easy}),
it is enough to know only 
the probability density  $\rho_\ep(\phi)$
of a single eigenphase. It is instructive to compare Eq.\ (\ref{easy})
with the similar representation of the integrated density of states
in the case of the normal-metal wire of finite length, which 
has been analysed recently.\cite{TMSB}

When the Andreev phase $\phi_A$ is real, i.e.,
for $\ep <\Delta$, the mean LDOS is found from Eq.\ (\ref{easy}) as  
\be
\label{elementary}
\bar{n}(x,\ep)=\pi \rho_\ep(\phi_A), \qquad \ep <\Delta,
\e
where the eigenphase density $\rho_\ep(\phi)$ is assumed to be
normalised to unity on the interval $(0,\pi)$.
The probability density $\rho_\ep(\phi)$ acquires its simplest
form in the case $N\gg 1$ of a large number of channels\cite{TB},
\be
\label{Ngg1}
\rho_\ep(\phi)
=\frac{1}{\pi \sin^2{\phi}}\im 
\sqrt{(\ep \tau_s)^2+i \ep \tau_s(1-e^{-2i\phi})},
\e
and in the single-channel case\cite{BG,WSZP},
\be
\label{Neq1}
\rho_\ep(\phi)=\frac{\ep \tau_s}{\pi}
\il_0^\infty \frac{\exp{(-\ep \tau_s t)}}{ t^2 \sin^2{\phi}
-t \sin 2\phi+1}dt.
\e

The scattering time $\tau_s$ of the DMPK scaling equation
differs by a numerical factor
(dependent on the dimensionality $d$
of the Fermi surface) from the mean scattering time of the
transport theory $\tau_s'$.
Namely, $\tau_s=c_d\tau_s^\prime$, 
where $c_d=2, \pi^2/4, 8/3$, 
for the dimensionality $d=1, 2, 3$, correspondingly.
%Namely, $\tau_s=\ell/v_F=c_d\tau_s^\prime=c_d^\prime \ell'/v_F$, 
%where $\ell'$ is the transport mean free path and
%$c_d=2, \pi^2/4, 8/3$, $c_d^\prime=1, \pi/2, 2$ 
%for the dimensionality $d=1, 2, 3$, correspondingly.
 
Note, that the integrated density of states (DOS)
$\nu = L^{-1}\int_0^L \bar{n}(x,\ep)\,dx$ in the infinite 
disordered wire $L\to\infty$ is given by the 
relation\cite{TBFM} $\nu=\pi (\pa /\pa\ep)
[\ep \rho_{\ep}(0)]$, which is similar 
in spirit to Eq.\ (\ref{elementary}). For wires with 
on-site disorder (in standard universality classes)
the value of $\pi \rho_{\ep}(0)$ equals to unity irrespective of energy 
$\ep$; however it can have a singularity at $\ep=0$ for wires with a
specific disorder symmetry. 

So far we were only concerned with the mean LDOS
for $\ep<\Delta$. However, the result (\ref{elementary})
can be easily extended to the energies
above the pair potential value 
with the help of the analytical continuation  
$\ep=i \omega$. On the other hand
the analytical continuation has another crucial
advantage. It transforms the dynamical correlator
$r_0(\ep)r_0(-\ep)^\ast$ into the essentially 
static object $r_0(i\omega)r_0(i\omega)^\ast$.
In the absence of a magnetic field the time-reversal symmetry
is preserved and the reflection matrix $r_0$
is symmetric, hence $r_0(i\omega)^\ast=r_0(i\omega)^\dagger$.
The eigenvalues $\exp(2 i\phi_j)$ of the matrix 
$r_0(\ep)r_0(-\ep)^\ast$ are transformed to
the real eigenvalues $R_j$ of the matrix
$r_0(i\omega)r_0(i\omega)^\ast$, which are
the probabilities of the reflection from
the long disordered wire in the presence of
a spatially uniform fictitious absorption $\omega$. 

The summation in Eq.\ (\ref{meso2}) is performed in 
terms of the eigenvalues 
\be
\label{easy2}
n(x,\ep)=\frac{1}{N}\re \lt. \s_{j=1}^N 
\frac{1-R_j \alpha^2(\omega)}{ 1+R_j \alpha^2(\omega)}
\rt|_{\omega=-i \ep +0^+},
\e
where $0^+$ is an infinitesimally small positive imaginary part 
of energy which ensures the retarded Green function 
required in Eq.\ (\ref{def_nu_mic}).  
We have also introduced  
\be
\label{alpha}
\alpha(\omega)=i e^{-i\phi_A}= 
\sqrt{1+(\omega/\Delta)^2}-\omega/\Delta.
\e

The joint probability density of the eigenvalues $R_j$ for
the infinitely long wire is given by the 
stationary solution of the DMPK equation.
In the parametrisation
\be
\label{par2}
R_j=\frac{\sigma_j}{ \sigma_j+2(N+1)\omega \tau_s},\quad
\sigma_j\in (0,\infty),
\e
this solution takes the simple form 
\be
\label{Laguerre}
P(\{\sigma_j \})=c_N
\p_{j=1}^Ne^{- \sigma_j/4}\p_{k>j}|\sigma_k-\sigma_j|,
\e
which we recognise as the 
orthogonal Laguerre ensemble of random matrix theory\cite{M}
(with normalisation constant $c_N$). This ensemble corresponds
to the class $CI$ in the classification scheme of Ref.\ \onlinecite{AZ}. 
The probability density (one-point function)
$\rho(\sigma)$, normalised to unity in the interval $(0,\infty)$,
is given by\cite{SN}
\beq
\hspace*{-1cm}
&&\rho(\sigma)=\frac{e^{-\sigma}}{ N}
\lt( \s_{n=0}^{N-1} \big[L_{n}^{(0)}(\sigma)\big]^2
-\frac{1}{ 2}L_{N-1}^{(0)}(\sigma)L_{N-1}^{(1)}(\sigma)\rt.\n \\
\label{orthogonal}
\hspace*{-1cm}
&&\lt. \qquad + \frac{1}{4} L_{N-1}^{(1)}(\sigma)\int_0^\sigma
d\zeta\, e^{(\sigma-\zeta)/2}L_{N-1}^{(0)}(\zeta)
\rt),
\eq
where $L_n^{(p)}(\sigma)$ is the generalised Laguerre polynomial.

We substitute the parametrisation (\ref{par2})
in Eq.\ (\ref{easy2}) and average over disorder with the help of 
the density $P(\{\sigma \})$. The result reads
\be
\label{result0}
\bar{n}(x,\ep)=\re \lt.\!\!\il_0^\infty \!d\sigma\,
\rho(\sigma)\frac{1-\frac{\alpha^2(\omega)-1}{2(N+1)\omega \tau_s}\,\sigma}
{1+\frac{\alpha^2(\omega)+1}{2(N+1)\omega \tau_s}\,\sigma}
\rt|_{\omega=-i\ep+0^+}.
\e
This equation extends Eq.\ (\ref{elementary}) to 
energies larger than $\Delta$. 
It can also be applied for arbitrary $N$.
In the large-$N$ limit
the distribution $\rho(\sigma)$ can be approximated by\cite{SN}
\be
\label{square}
\lim_{N\to\infty}N\rho(\zeta N)=\frac{1}{ 2\pi}\sqrt{\frac{4}{\zeta}-1}.
\qquad 0<\zeta < 4.
\e
Substituting this expression into
Eq.\ (\ref{result0}) we
reproduce the results of
Eqs.\ (\ref{elementary},\ref{Ngg1}) 
for $\ep < \Delta$.
In the limit $\ep\to 0$ this 
leads to the square root behavior of the LDOS
$\bar{n}(x,\ep\to 0)=\re \sqrt{-i \ep\tau_s}$.
In the extremely dirty regime $\Delta\tau_s\to 0$
we reproduce the result of the conventional BCS theory
\be
\label{BCST1}
\bar{n}(x,\ep)=\re \ep/\sqrt{\ep^2-\Delta^2}.
\e

The Thouless energy $\ep_c=1/N^2\tau_s$, however,
remains unresolved within the multi-channel approximation (\ref{square}). 
In order to fix the scale $\ep_c$ one has to
take advantage of another limiting relation\cite{NS}
\beq
&&\lim_{N\to\infty}\rho(\zeta/N)=
J_1^2(2\sqrt{\zeta})-J_0(2\sqrt{\zeta})J_2(2\sqrt{\zeta})\n\\
&&\qquad
+\lt( 2\sqrt{\zeta}\rt)^{-1} J_0(2\sqrt{\zeta})J_1(2\sqrt{\zeta}),
\label{obscure}
\eq
where $J_n(z)$ are Bessel functions.
In the limit $\ep\to 0$ one can safely
put $\alpha(\omega)=1$ in Eq.\ (\ref{result0})
and take the real part explicitly,
\be
\label{add_to_one}
\bar{n}(x,\ep)=\il_0^\infty d\sigma\,\rho(\sigma)
\frac{1}{1+\sigma^2 [(N+1)\ep \tau_s]^{-1}}.
\e
To leading order in $\ep/\ep_c$ 
the function $\rho(\sigma)$ in (\ref{add_to_one}) 
can be approximated 
by its value at the origin $\rho(0)=1/2$, which holds for 
$\sigma \ll N^{-1}$ [see Eq.\ (\ref{obscure})]. We therefore obtain
\be
\label{loc_limit}
\bar{n}(x,\ep)=\frac{\pi (N+1)}{4} \ep \tau_s
\approx \frac{\pi}{4 N} (\ep/\ep_c), \qquad \ep \ll \ep_c.
\e
The factor $1/N$ in the last expression reflects the fact
that only a single channel is responsible for the non-vanishing 
LDOS at energies lower than $\ep_c$. 

In Fig.~\ref{fig:figT1}
we plot the mean LDOS given by Eq.\ (\ref{result0})
against the ratio $\ep/\Delta$
for different numbers of channels in the moderately
dirty regime $\Delta \tau_s = 0.3$, 
the intermediate regime $\Delta\tau_s =1$, 
and the moderately clean regime $\Delta \tau_s=3$.
We observe that the curves are lying close 
to each other in all cases. (This suggests that
the LDOS near the ideally transmitting interface is
quite insensitive to phase-coherent effects.)
The situation changes in the case of a finite transparency
$\Gamma <1$ of the NS interface. 

\section{Effect of a tunnel barrier} \label{sc:Effect}

\subsection{Model}

We now introduce the simplest model of a dielectric
tunnel barrier at the ideal NS interface. 
The mean LDOS is calculated 
in the normal-metal
at a ballistic distance $x \ll \ell$ from the interface
(see Fig.~\ref{fig:setup}). 

We describe the segment $I$ of the wire 
between the chosen cross section 
and the ideal NS interface 
(this segment includes the tunnel barrier)
by its $S$-matrix 
\be
\label{SI}
S_I=
\lt(\ba{cc}
r_1^I& t_2^I \\ t_1^I & r_2^I
\ea 
\rt),
\e
where each block itself consists of block-diagonal 
matrices in the particle-hole representation,
\beml
\label{rItI}
\beq
r_{1,2}^I &=&
\lt(\ba{cc}
r_{1,2}(\ep)& 0 \\ 0 & r_{1,2}(-\ep)^\ast
\ea
\rt), \\
t_{1,2}^I &=&
\lt(\ba{cc}
t_{1,2}(\ep)& 0 \\ 0 & t_{1,2}(-\ep)^\ast
\ea\rt),
\eq
\eml
and the matrices $r_{1,2}(\ep)$, $t_{1,2}(\ep)$ are $N\times N$
electron reflection and transmission matrices corresponding to
the segment $I$.

The right matrix $r_R$ in the fundamental formula (\ref{meso2})
depends on the $S$-matrix of the segment $I$ 
[see Eqs.\ (\ref{SI},\ref{rItI})] 
and on the scattering matrix for Andreev reflection 
[see Eq.\ (\ref{par1})]. 
A straightforward 
algebraic calculation gives\cite{CWJ}
\beml
\label{combined}
\beq
\hspace*{-1cm}
{r}_R&=&
\lt(\ba{cc}
r_c(\ep)& -t_c(-\ep)^\ast \\ 
t_c(\ep)& r_c(-\ep)^\ast
\ea
\rt),\\
\hspace*{-1cm}
t_c(\ep)&=& e^{-i\phi_A(\ep)} t_2(-\ep)^\ast M(\ep) t_1(\ep),\\
\hspace*{-1cm}
r_c(\ep)&=&r_1(\ep)+e^{-2 i \phi_A(\ep)}t_2(\ep)r_2(-\ep)^\ast
M(\ep)t_1(\ep),\\
\hspace*{-1cm}
M(\ep)&=&\lt[1-e^{-2 i\phi_A(\ep)}r_2(\ep) r_2(-\ep)^\ast\rt]^{-1}.
\eq
\eml
 
In general, if the segment $I$ contains some weak disorder
(which is the case, for example, for $x>\ell$) 
the correlations between the matrices 
$r_{1,2}$ and $t_{1,2}$ 
for electron- and hole-like quasiparticles 
are non-trivial. We consider here the case
that the segment $I$ contains no disorder, but
a sufficiently steep tunnel barrier
which makes no difference in the tunnelling probability 
of electrons and holes. In this case we can omit the energy 
dependence in the matrices
$r_{1,2}$ and $t_{1,2}$.
In what follows we take advantage of the polar decomposition
\be
\label{decompose1}
\lt(\ba{cc}
\!r_1\! & \!t_2\!\\ \!t_1\! & \!r_2\!
\ea\rt)=\lt(\ba{cc}
\!u_I\! & \!0\! \\ \!0\! & v_I^T\!\!
\ea\rt)
\lt(\ba{cc}
\!\sqrt{1\!-\!\Gamma}\! & \!i\sqrt{\Gamma}\!\\
\!i\sqrt{\Gamma}\! & \!\sqrt{1\!-\!\Gamma}\!
\ea\rt)
\lt(\ba{cc}
\!u_I^T\! & \!0\! \\ \!0\! & \!v_I\!
\ea\rt),
\e
where $u_I$, $v_I$ are some unitary matrices, which depend
on a particular realisation of the barrier, and
$\Gamma$ is the diagonal matrix of the tunnelling
probabilities $\Gamma_j$. Time-reversal symmetry in 
the segment $I$ is assumed.
Once the dependence on energy in the matrices 
$u_I$, $v_I$ and $\Gamma$ is disregarded
we obtain from Eqs.\ (\ref{combined},\ref{decompose1})
the right reflection matrix 
\beml
\label{barrier}
\beq
\label{rR}
\hspace*{-1cm}
&&{r}_R
=\!\lt(\ba{cc}
\!u_I\! & \!0\! \\ \!0\! & \!u_I^\ast\!
\ea\rt)\!\!
\lt(\ba{cc}
\!e^{i\chi}\cos\theta\! & \!-ie^{i\chi}\sin\theta\!\\
\!-ie^{i\chi}\sin\theta\! & \!e^{i\chi}\cos\theta\!
\ea\rt)\!\!
\lt(\ba{cc}
\!u_I^T\! & \!0\! \\ \!0\! & \!u_I^\dagger\!
\ea\rt)\!,\\ \label{theta}
\hspace*{-1cm}
&&\sin{\theta}=
\Gamma\lt[(1\!-\!e^{2i\phi_A}(1\!-\!\Gamma))
(1\!-\!e^{-2i\phi_A}(1\!-\!\Gamma))\rt]^{-\frac{1}{2}}\!,\\
\label{chi}
\hspace*{-1cm}
&&\mbox{ }
e^{2i\chi}=(1-\Gamma-e^{2i\phi_A})\lt[1-e^{2i\phi_a}(1-\Gamma)\rt]^{-1}.
\eq
\eml

The left matrix $r_L$ is given by Eq.\ (\ref{par1})
and describes the reflection from the disordered wire.
Taking advantage of the polar decomposition we can write
\be
\label{rL}
{r}_L=
\lt(\ba{cc}
u_0 & 0 \\ 0 & u_0^\ast
\ea\rt)
\lt(\ba{cc}
e^{i\phi} & 0 \\
0 & e^{i\phi}
\ea\rt)
\lt(\ba{cc}
u_0^T & 0 \\ 0 & u_0^\dagger
\ea\rt),
\e 
where $u_0$ is a random unitary matrix and 
$\phi$ is the diagonal matrix of the eigenphases.
We see that all information contained in $u_I$
disappears statistically from the eigenvalues of $r_L r_R$
because the product $u_I^T u_0$ can be regarded again as 
a random unitary matrix. Thus 
the disorder-averaged
LDOS depends only on the transmission eigenvalues 
$\Gamma_j$ of the tunnel barrier. 
Below we calculate the mean LDOS
for a single-channel wire and for a multi-channel wire
provided the tunnelling probabilities are the same for
all channels, i.e., $\Gamma_j=\Gamma$.  

%%%%%%%%%%%%%%%%%%%%%%%%%%%%%%%%%
\begin{figure}
\includegraphics[width=7.5cm]{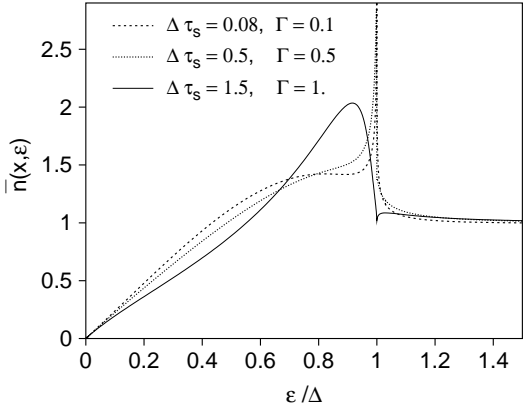}
\caption{
The mean LDOS (\ref{1modeLDOS}) in a single-channel
disordered wire at ballistic distance 
from an NS interface of finite transparency.  
The parameters $\Delta\tau_s$ and $\Gamma=0.5$
are chosen to fix the combination 
$\Delta \tau_s (2-\Gamma)/\Gamma\approx 1.5$. 
The curves are close to each other
for $\ep \ll \Delta$ where Eq.\ (\ref{T1ellDelta})
is applicable.
}
\label{fig:figSc1}
\end{figure}
%%%%%%%%%%%%%%%%%%%%%%%%%%%%%%%%%

\subsection{Single channel wire}

We start with the calculation of the mean
LDOS for $\ep<\Delta$ in the case of the
single-channel wire $N=1$.
For  $\ep<\Delta$ the phases $\chi$ and $\theta$ 
defined in Eq.\ (\ref{barrier}) are real
and both $r_L$ and $r_R$ are unitary $2\times 2$ matrices. 
We denote 
$u_I^Tu_0=\exp(i\psi)$, where $\psi$ is a random
phase distributed uniformly in the interval $(0,2\pi)$. 
We insert the reflection matrices from
Eqs. (\ref{rR},\ref{rL}) directly to Eq.\ (\ref{meso}).
The matrix $(1-r_L r_R)$ can be easily inverted.
Taking the real part we notice that 
the zeroes of $\det(1-r_L r_R)$ define the
exact positions of the quasiparticle
bound states for $\ep <\Delta$. The result reads
\be
\label{Neq1T}
n(x,\ep)=\pi \sin(\phi+\chi)\
\delta\!\lt(\cos\theta\cos\psi-\cos(\phi+\chi)\rt), 
\e
where the argument of the Dirac $\delta$-function  
corresponds to the quantisation condition for the
bound states. The mean LDOS is given by the average 
over the phase $\phi$
with the probability density $\rho_{\ep}(\phi)$ of Eq.\ (\ref{Neq1}),
and over the uniformly distributed phase $\psi$.
The integration over $\psi$ is readily done with the result
\be
\label{Tsubgap}
\bar{n}(x,\ep)=\!\!\!\il_{\theta-\chi}^{\pi-\theta-\chi}
\!\!\!d\phi\ \rho_{\ep}(\phi)\frac{\sin(\phi+\chi)}{ 
\sqrt{\cos^2{\theta}-\cos^2(\phi+\chi)}}.
\e

In the limit $\Gamma \to 1$
of the vanishing tunnel barrier one
observes that $\chi \to \pi/2-\phi_A$ and $\theta\to \pi/2$,
so that the area of the integration in Eq.\ (\ref{Neq1T})
shrinks to the small vicinity of $\phi=\phi_A$ and the function
$\rho_{\ep}(\phi)$ can be substituted by its value in 
this point. The integral approaches $\pi$ and
we recover the result of Eq.\ (\ref{elementary})
for the ideally transmitting interface.

In the opposite extreme of a high tunnel barrier ($\Gamma\to 0$)
both $\theta$ and $\chi$ go to zero, so that the integration
area is not restricted and the value of the integral 
tends to unity because of the normalisation condition for 
the probability density $\rho_{\ep}(\phi)$.
In the limit $\ep \ll \Delta$ we can set $\chi=0$ 
and reduce Eq.\ (\ref{Tsubgap}) to the following form
\be
\label{T1ellDelta}
\bar{n}(x,\ep)=\!\!\re\! \il_0^\infty\! dt\, e^{-t}
\lt[1\!-\!\frac{\sin^2\theta}{ (\ep \tau_s)^2}\,  t\,
(t-2 i \ep\tau_s) \rt]^{-\frac{1}{2}},
\e
where  $\sin\theta =\Gamma/(2-\Gamma)$, according to Eq.\ (\ref{theta}).
From Eq.\ (\ref{T1ellDelta}) we find that 
\be
\bar{n}(x,\ep)=\pi \ep \tau_s 
\frac{2-\Gamma}{2\Gamma}, \qquad 
\ep \tau_s \ll \frac{\Gamma}{2-\Gamma},
\e
which coincides for $\Gamma=1$ with the result of
Eq.\ (\ref{loc_limit}) for $N=1$.
In the dirty limit $\Delta \ll \tau_s^{-1}$ and for a high
tunnel barrier $\Gamma \ll 1$ the result 
of Eq.\ (\ref{T1ellDelta}) is applicable
almost up to the value of $\ep=\Delta$.
It describes the formation of the pseudo-gap 
near the  energy $\tau_s^{-1}\Gamma$ 
due to the normal reflection from the barrier.
 
The exact expression  (\ref{Tsubgap}) 
additionally accounts for 
the peak at $\ep \simeq \Delta$.
This expression can be further generalised for energies
higher than $\Delta$ by means of the analytical continuation
$\ep=i\omega$, with the result
\beml
\label{1modeLDOS}
\beq
\hspace*{-1cm}
\bar{n}(x,\ep)&=&-\re\il_0^\infty d\sigma\
\rho_1(\sigma)\frac{\sinh{Q(\sigma)}}{
\sqrt{\sinh^2{Q(\sigma)}+\sin^2{\theta}}},\\
\hspace*{-1cm}
Q(\sigma)&=&\frac{1}{2}\ln\frac{\alpha^2(\omega)+1-
\Gamma}{1+\alpha^2(\omega)(1-\Gamma)}+
\frac{1}{2}\ln\frac{\sigma}{\sigma+4\omega\tau_s},
\eq
\eml
where the function
$\rho_1(\sigma)=(1/2)\exp(-\sigma/2)$ is the probability 
density (\ref{orthogonal}) for a single-channel 
wire, the function $\alpha(\omega)$ is defined in Eq.\ (\ref{alpha}),
and the continuation to the real energies
$\omega\to -i\ep +0^+$ is performed.

\subsection{Multi-channel wire}

The disorder-averaged LDOS for $N\gg 1$
can be found straightforwardly
for the case of
equivalent tunnelling probabilities $\Gamma_j=\Gamma$. 
Then the diagonal matrices $\theta$ and 
$\chi$ in Eqs. (\ref{theta},\ref{chi}) 
can be regarded as scalars. 
It is convenient to make use of the analytical continuation
$\ep=i \omega$ and define
\beml
\label{definitions}
\beq
\hspace*{-1cm}
&&\alpha(\omega)=ie^{i \phi_A}, \qquad
R=e^{2i\phi},
\\
\hspace*{-1cm}
&&
p=e^{2i\chi}=\frac{\alpha^2(\omega)+1-\Gamma}
{1+\alpha^2(\omega)(1-\Gamma)}, 
\eq
\eml
where $R=\diag(R_1, \dots R_N)$ is the diagonal matrix
of reflection probabilities for the disordered wire
with a fictitious absorption $\omega$.
In the parametrisation (\ref{par2})
the joint probability density of $R_j$ is related to the 
orthogonal Laguerre ensemble (\ref{Laguerre}).
Note that the quantities $p$, $R_j$ and $\alpha(\omega)$
take real values in the interval $(0,1)$
when $\omega$ is real. 

The basic expression (\ref{meso2}) for the mean LDOS
is manifestly invariant under an arbitrary unitary
rotation of the matrix product $r_L r_R$. From 
Eqs.\ (\ref{rR},\ref{rL}) we obtain
\beml
\beq
\label{product}
\hspace*{-1cm}
&&U_0^\dagger
{r}_L {r}_R 
U_0
=\!
\lt(\ba{cc}
\!\!\sqrt{pR}\! & \!\!0\!\\ \!0\! & \!\sqrt{pR}\!
\ea
\rt)\!\! 
\lt(\ba{cc}
\cos\theta\ U & -i\sin\theta\\ 
-i\sin\theta & \cos\theta\ U^\dagger
\ea
\rt),\\ \label{construction}
\hspace*{-1cm}
&&
U=u_0^T u_I u_I^T u_0,
\qquad 
U_0=\diag(u_0,u_0^\ast),
\eq
\eml
where we take advantage of the quantities defined in
Eq.\ (\ref{definitions}). The matrix $u_0$ is a random
unitary matrix which is uniformly distributed 
in the unitary group (provided the weak disorder 
$k_F \ell \gg 1$). Hence by construction (\ref{construction}),
$U$ is the unitary symmetric random matrix.
We substitute Eq.\ (\ref{product}) into Eq.\ (\ref{meso2}) 
to express the mean LDOS as
\be
\label{meanLDOS}
\bar{n}(x,\ep)=\frac{1}{N}\re \tr \lt\la 
\frac{1-p R}{1+p R}\ \Big\la{\!F(\cos\theta)}
\Big\ra_{\!U}
\rt\ra_{\!\!\!R},
\e
where 
\beml
\label{F(z)}
\beq
\hspace*{-1cm}
&&F(z)=\frac{1}{1-z 
(\sqrt{C_1}\, U \sqrt{C_2}+\sqrt{C_2}\, U^\dagger \sqrt{C_1})},\\
\hspace*{-1cm}
&&
\label{C1C2}
C_1=\frac{p R}{1+p R}, \qquad C_2=\frac{1}{1+p R}.
\eq
\eml
The average over disorder in Eq.\ (\ref{meanLDOS})
is decoupled into two independent steps: 
the average $\la\dots \ra_U$ over the group spanned by
the unitary symmetric matrices
and the average $\la \dots \ra_R$ over the 
orthogonal Laguerre ensemble of the reflection eigenvalues $R_j$.

In the case of the finite number of channels 
the calculation of average over the unitary matrices $U$
is technically difficult and cannot be done analytically. 
However, for the diffusive wire, $N\gg 1$, 
the calculation can be done
by means of the diagrammatic 
technique developed in Ref.~\onlinecite{BB}.

Let us briefly quote the basic substitution
rules of the diagrammatic technique 
\be\ba{rclcrcl}
U_{ij}&=&\includegraphics[width=1.2cm]{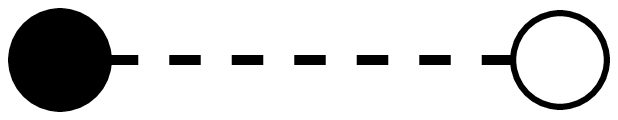},
&\qquad&
U_{ij}^\ast&=&\includegraphics[width=1.2cm]{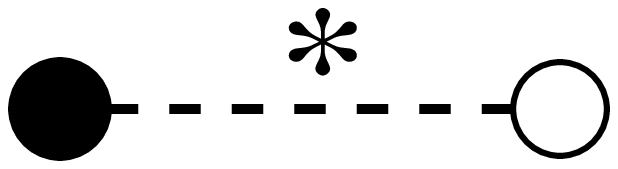},\\
C_{ij}&=&\includegraphics[width=1.2cm]{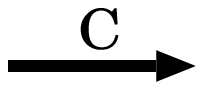},
&\qquad&
\delta_{ij}&=&\includegraphics[width=1.2cm]{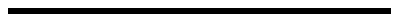}.
\ea
\e
Here the matrix element $U_{ij}$ 
is represented by the black and white dot
connected by the dashed line. The black dot stays for the first 
index $i$ and the white dot for the second index $j$.
The conjugated matrix $U^*$ is marked by a star.
The other matrices are denoted by thick solid  
arrows. The summation over a matrix 
index in a dot is indicated by the attachment of a 
solid line. The average over the unitary symmetric matrices is 
symbolically performed by pairing 
in all possible ways
all black and white dots belonging to $U$ 
to all black and white dots 
belonging to $U^\ast$. 
This pairing is denoted by the thin solid line, 
which corresponds to
the Kronecker symbol. The result of 
the averaging 
is found by inspection of the closed circuits 
in the diagram which consist of alternating thick and thin solid
lines (T-circles). Each diagram is weighted by a
factor, which is obtained by 
inspection of the closed circuits 
of alternating thin solid and dashed lines (U-circles). 

\begin{figure}
\includegraphics[width=7.5cm]{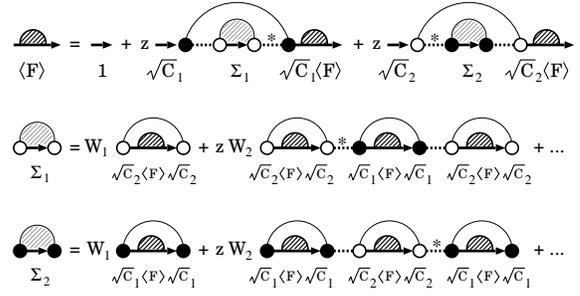}
\caption{
Diagrammatic representation of the Dyson equation
(\ref{DysonO}) for $\la F(z)\ra_{U}$.
}
\label{fig:figDyson}
\end{figure}

We expand the matrix $F(z)$ (\ref{F(z)})
into a geometric series and keep only the terms with equal number 
of $U$ and $U^\dagger$ matrices. In the large-$N$ limit 
we have to take into account the diagrams 
with the largest number of T-circles\cite{BB}. 
This amounts to the summation of the
`rainbow' diagrams, or diffusion ladders, depicted 
symbolically in Fig.~\ref{fig:figDyson}. The corresponding
Dyson equation is
\beml
\label{DysonO}
\beq
\hspace*{-1cm}
&&\la F\ra_U=
\hat{1}+z \Sigma_1 C_1 
\la F \ra_{U}
+ z \Sigma_2 C_2 
\la F\ra_{U},\\ 
\hspace*{-1cm}
&&\Sigma_1=\s_{n=1}^\infty
W_n z^{n-1}\lt[\tr C_2 \la F\ra_U\rt]^n 
\lt[\tr C_1 \la F\ra_U\rt]^{n-1},\\ 
\hspace*{-1cm}
&&\Sigma_2=\s_{n=1}^\infty
W_n z^{n-1}\lt[\tr C_1 \la F\ra_U\rt]^n 
\lt[\tr C_2 \la F\ra_U\rt]^{n-1},
\eq
\eml
where the weight factors
\be
\label{weights}
W_n=N^{1-2 n}(-1)^{n-1}\frac{(2n-2)!}{n!(n-1)!}
+{\cal O}\lt(N^{-2 n}\rt) 
\e
have been found in Ref.\ \onlinecite{BB}. Taking the
coefficients $W_n$ to the leading order in $N$ 
we define the generating function 
\be
h(s)=\s_{n=1}^\infty
W_n s^{n-1}=\frac{1}{2 s}\lt(
\sqrt{N^2+4s}-N\rt),
\e
which may be used to reduce Eq.\ (\ref{DysonO}) to
\beq
\big\la\!F(z)\big\ra_{U}&=&\hat{1}+z^2 h(z^2 s_1 s_2)\
(s_1 C_1+s_2 C_2)\ \big\la\!F(z)\big\ra_{U},\n \\
\label{DysonO1}
s_{1,2}&=&\tr C_{1,2}\,\big\la\!F(z)\big\ra_{U}.
\eq

%%%%%%%%%%%%%%%%%%%%%%%%%%%%%%%%%%%%%%%%%
%%%%%%%%%%%%%%%%%%%%%%%%%%%%%%%%%%%%%%%%%%

The matrix $\la F(z) \ra_U$
has to be eliminated from the Dyson equations (\ref{DysonO1}).
After that it is very convenient to transform to the new 
scalar variables
\be
\label{scalar}
X=\frac{s_2-s_1}{N}, \qquad Y=\frac{s_2+s_1}{N},
\e
which obey the equations 
\beml
\label{DysonO3}
\beq
\label{justX}
&&\frac{X+1}{2}=\frac{1}{N} \tr \frac{1}{
1+p R\, f(X,Y)},\\
&&Y^2\sin^2\theta+ X^2 \cos^2\theta=1,
\eq
\eml
with
\be
f(X,Y)=\frac{(1-X)(Y+X)}{(1+X)(Y-X)},
\e
where we have substituted $z=\cos{\theta}$
and the matrices $C_1$, $C_2$ from Eq.\ (\ref{C1C2}).

In terms of the variables $X$ and $Y$
the mean LDOS (\ref{meanLDOS}) is simplified to
\be
\label{exciting}
\bar{n}(x,\ep)=\re \lt.\overline{X}(\omega)\rt|_{\omega\to -i\ep+0^+}, 
\e
where the bar stands for the 
average over the ensemble
of the reflection probabilities: $\overline{X}\equiv \la X \ra_R$.  

Let us first consider the case  
of equal reflection probabilities $R_j=R$.
The matrix $U$ in Eq.\ (\ref{F(z)})
commutes with $C_1$ and $C_2$ 
and can be diagonalised,
hence the problem becomes equivalent to 
that of a single channel wire.
The solution of the self-consistent 
equations Eq.\ (\ref{DysonO3}) is given by
\be
\label{limiting}
X=-\frac{\sinh{Q}}{\sqrt{\sinh^2{Q}
+\sin^2{\theta}}},\quad Q=\frac{1}{2}\ln p R,
\e
which coincides with the result of the exact 
integration over $U$. This proves that
the set of diagrams which we took into account in
Eqs.\ (\ref{DysonO}) is sufficiently complete.
The substitution of Eq.\ (\ref{limiting}) 
in  Eq.\ (\ref{exciting}) and the additional average
over the reflection probability of a single channel wire
yields the mean LDOS of Eq.\ (\ref{1modeLDOS}).

In the multichannel (diffusive) limit $N\gg 1$
the reflection probabilities $R_j$ are, in fact,  not 
equal. Moreover they effectively repel each other according to 
Eqs.\ (\ref{par2},\ref{Laguerre}).
In this case Eq.\ (\ref{DysonO3}) can no longer be solved 
in closed form. In other words, the averages 
over the random matrix $U$ and 
over the reflection eigenvalues $R_j$ cannot be 
performed separately.

%%%%%%%%%%%%%%%%%%%%%%%%%%%%%%%
%%%%%%%%%%%%%%%%%%%%%%%%%%%%%%%%%

In order to proceed one has to take advantage of
the self-averaging property of 
the variables  $X$ and $Y$ in the limit $N\gg 1$.  
Indeed both variables are defined via 
the traces $s_{1,2}$ and can be thought 
as the arithmetic means of $N$ fluctuating quantities.
From physical point of view the variable
$X$ is proportional to the one-point Green function, 
therefore it is self-averaging in a diffusive metal.

%%%%%%%%%%%%%%%%%%%%%%%%%%%%%%%
\begin{figure}
\includegraphics[width=7.5cm]{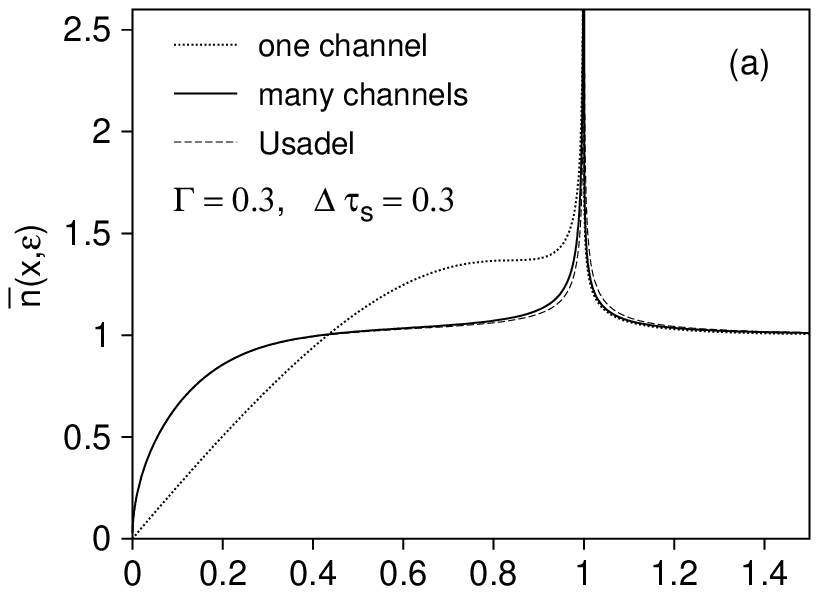}
\includegraphics[width=7.5cm]{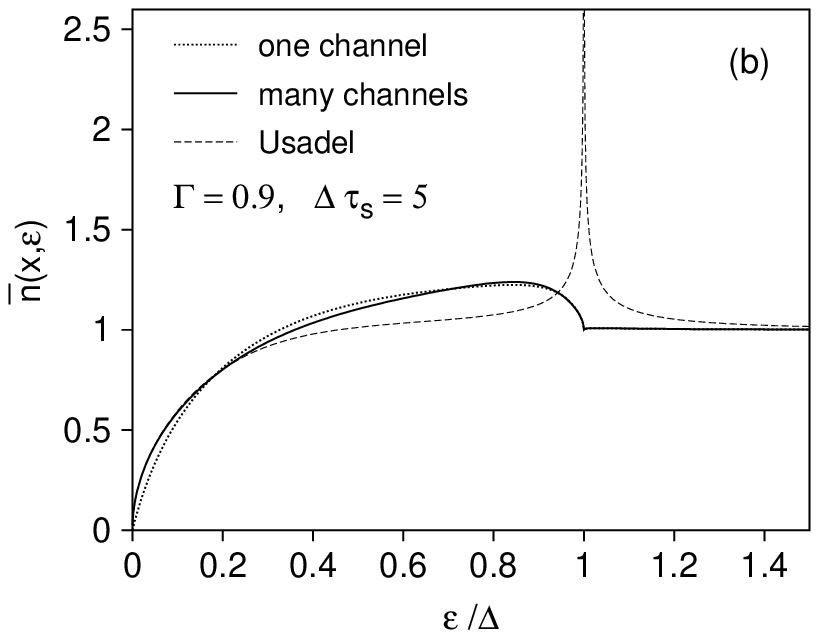}
\caption{
The mean LDOS in a normal-metal wire
in the vicinity of an NS interface
of finite transparency. 
The dotted curve is found from Eq.\ (\ref{1modeLDOS})
for the single-channel wire $N\!=\!1$. The solid curve
is calculated from Eqs.\ (\ref{exciting},\ref{InftymodeLDOS})
for the diffusive wire $N\!\gg\!1$. 
The dashed curve represents the result of the Usadel equation, 
Eqs.\ (\ref{Uldos1},\ref{Uresult}), calculated for the
corresponding value of the parameter 
$\gamma_B^2=\Delta\tau_s(2-\Gamma)^2/(2\Gamma)^2$.
The figures show the energy dependence of the 
mean LDOS for the dirty $\Delta \tau_s=0.3$ (top)
and the clean $\Delta\tau_s=5$ regime.
}
\label{fig:figTa}
\end{figure}
%%%%%%%%%%%%%%%%%%%%%%%%%%%%%%%%%

Thus we can construct the self-consistent 
equation for $\overline{X}$ by
taking the average over $R$
on both sides of Eq.\ (\ref{justX}).
We assume a fixed value of $f(X,Y(X))=
\tilde{f}(\overline{X})$
on the right side, 
neglecting the fluctuations of $X$. 
Taking advantage of the
square-root approximation (\ref{square})
of the density $\rho(\sigma)$
we obtain
\be
\frac{\overline{X}+1}{2}=
\frac{1}{2\pi}\!\il_0^4\!\! d\zeta \sqrt{\frac{4}{\zeta}-1}\
\frac{2\omega \tau_s+\zeta}
{2\omega\tau_s\! +\!\lt(1\!+\!p \tilde{f}(\overline{X})\rt) \zeta}.
\e
The integral on the right-hand side
can be carried out explicitly giving rise to the 
equation
\beq
\hspace*{-1cm}
&&\frac{\lt(\alpha^2(\omega)+1-\Gamma\rt)\lt(Y\lt(\overline{X}\rt)+\overline{X}\rt)
}{\lt(1+ \alpha^2(\omega)(1-\Gamma)\rt)\lt(
Y\lt(\overline{X}\rt)-\overline{X}\rt)}\n\\
\label{InftymodeLDOS}
\hspace*{-1cm}
&&=1+\frac{2}{ 1+\overline{X}}
\lt(\omega\tau_s -\sqrt{\omega\tau_s}\sqrt{1+
\overline{X}+\omega\tau_s}\rt),
\eq 
which is an algebraic equation for $\overline{X}$.
It can be analytically continued to real energies
$\omega =-i\ep+0^+$ and solved numerically by iteration.
The disorder-averaged LDOS is determined, then, from 
Eq.\ (\ref{exciting}).
Equation (\ref{InftymodeLDOS}) is obtained in the quasiclassical limit of 
a large number of channels. This result does not change
if we neglect that $U$ is symmetric or
take the unitary Laguerre ensemble in Eq.\ (\ref{Laguerre})
instead of the orthogonal one.

The weak-localisation correction 
(which we simply define as $1/N$ correction)
can, in principle, be determined 
within the present approach. It has three different sources. 
First of all an additional class
of diagrams, namely the cooperon-like diagrams, 
have to be taken into account in the Dyson equation (\ref{DysonO}).
Secondly the term of sub-leading order in the large-$N$
expansion of the weight factors $W_n$ has to be included.
Finally the correction of order ${\cal O}(N^{-1})$ to the limiting form  
(\ref{square}) of the probability density $\rho(\sigma)$
has to be considered.
The calculation of the weak localisation correction to the LDOS  
is, however, beyond the scope of this paper.

In some limiting cases Eq.\ (\ref{InftymodeLDOS}) allows for 
a transparent analytical solution. In the absence of the tunnel 
barrier $\Gamma=1$ we obtain
\be
\overline{X}=1-2 \frac{\alpha^2}{
1+\alpha^2} 
\lt.\lt(1+\Omega-\sqrt{\Omega}\sqrt{2+\Omega}\rt)
\rt|_{\Omega\to\frac{\omega\tau_s}{ 1+\alpha^2}},
\e
which coincides upon the substitution in (\ref{exciting})
with the result of Eq.\ (\ref{result0}) in the large-N limit.

In the limit $\Delta \tau_s\ll \Gamma^2$, 
Eq.\ (\ref{InftymodeLDOS}) leads to the BCS result for the 
local density, Eq.\ (\ref{BCST1}).

\begin{figure}
\includegraphics[width=7.5cm]{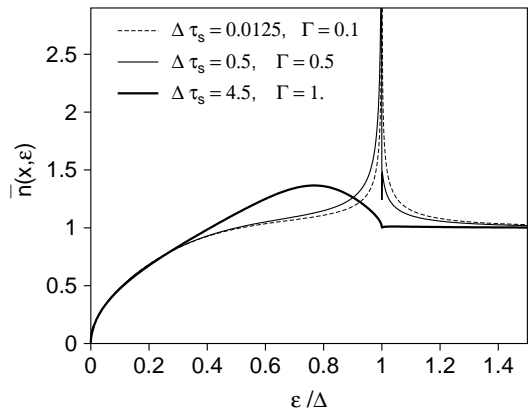}
\caption{
The mean LDOS in vicinity of an NS interface of finite transparency
calculated from Eqs.\ (\ref{exciting},\ref{InftymodeLDOS}).
The parameters $\Delta\tau_s$ and $\Gamma=0.1$
are chosen to fix the combination 
$\gamma_B^2=\Delta \tau_s (2-\Gamma)^2/(2\Gamma)^2
\approx 1.06$.
The curves coincide for $\ep \ll \Delta$, 
where Eqs.\ (\ref{scaling},\ref{Uscaling}) are applicable. 
The result of the Usadel equation, 
Eqs.\ (\ref{Uldos1},\ref{Uresult}), 
is indistinguishable from the dashed line. 
}
\label{fig:figScI}
\end{figure}

For small energies,
$\omega \ll \Delta$, we can put $\alpha(\omega)=1$
and obtain
\be
\overline{X}=\frac{\sqrt{\omega\tau_s}
\sqrt{4\sin^2\theta+\omega\tau_s}-\omega\tau_s
}{ 2 \sin^2\theta}.
\e
The mean LDOS (\ref{exciting}) for $\ep\ll \Delta$
is then given by
\be
\label{scaling}
\bar{n}(x,\ep)=\re \sqrt{-i \frac{\ep \tau_s}{\sin^2\theta}}
\sqrt{1-i\frac{\ep \tau_s}{4 \sin^2\theta} },
\e
with $\sin\theta=\Gamma/(2-\Gamma)$. 
This result describes the scaling 
$\ep_g\sim \tau_s^{-1}\Gamma^2(2-\Gamma)^{-2}$
of the size of the pseudogap $\ep_g$
with the transparency of the tunnel barrier
$\Gamma$,
which is illustrated in Fig.~\ref{fig:figScI}.
We observe that 
in the limit $\Gamma^2\ll \Delta\tau_s
\ll 1$ two different types of bound
states contribute to the LDOS 
at energies below $\Delta$. 
One group of the bound states is responsible 
for the monotonous increase 
of the LDOS to its bulk value at the scale 
$\tau_s^{-1}\Gamma^2$ while another group
gives rise to the formation of the peak 
near $\ep = \Delta$.

%%%%%%%%%%%%%%%%%%%%%%%%%%%%%%%
%%%%%%%%%%%%%%%%%%%%%%%%%%%%%%%%%

\section{Usadel equation} \label{sc:Usadel}

The aim of this Section is to compare our results 
in the limit $N\gg 1$ to the results of the conventional 
quasiclassical theory based on the Usadel equation. 
It is important to remember that
the Usadel description is justified only
in the dirty limit $\Delta\tau_s\ll 1$,
while it is not restricted to the clean superconducting material 
as is the case with our calculation. 
In the quasiclassical context the superconductor
as well as the normal metal are characterised by their 
diffusion constants $D_s$, $D_n$ and normal-state resistivities
$\rho_s$, $\rho_n$, which are combined into the mismatch parameter 
\be
\gamma=\frac{\rho_s \xi_s}{\rho_n \xi_n},
\e
where $\xi_{n,s}=\sqrt{D_{n,s}/\Delta}$ are
the diffusive coherence lengths. 
Hence, the comparison has to be done in the limit $\gamma\ll 1$,
where the ``rigid'' boundary condition is valid.

In the case of the perfectly transparent NS boundary
and vanishing mismatch parameter the LDOS at the interface
found from the Usadel equation\cite{GK88,BBS}
is simply given by the standard BCS formula
and, therefore, coincides with our expression (\ref{BCST1}) 
in the dirty limit $\Delta\tau_s\ll 1$.
Thus, there is not too much to compare for the case of
transparent boundary.
However, if the NS interface is not perfectly transparent
($\Gamma < 1$), even 
the limit of small mismatch parameter $\gamma$ 
is not completely trivial. 
Let us now discuss the Usadel equation for this case
in somewhat more detail following the calculation 
of Ref.\ \onlinecite{GK96}. 

The transparency of the interface enters the 
theory through the parameter 
\be
\gamma_B=\frac{R_B}{\rho_n\xi_n},
\e
where $R_B$ is the product of the barrier
resistance and its area. 
The Usadel equation in the normal metal ($x<0$)
takes the form
\be
\label{UsadelN}
\frac{D_n}{2}\Theta''_n(x)-\omega \sin\Theta_n(x) =0,
\e
where $\omega=-i \ep+0^+$ is the imaginary energy, while
in the superconductor ($x>0$) the equation reads
\be
\label{UsadelS}
\frac{D_s}{2}\Theta''_s(x)-\omega \sin\Theta_s(x)
+\Delta(x)\cos\Theta_s(x) =0,
\e
where $\Delta(x)$ is the gap function. 
(For a sake of simplicity we restrict ourselves 
to zero temperature.) The functions 
$G(x,x)=\cos\Theta_{n,s}(x)$ and $F(x,x)=\sin\Theta_{n,s}(x)$
parametrise normal and anomalous quasiclassical Green functions
in energy representation, averaged over angle and disorder.  
The LDOS near the interface is given by
\be
\label{Uldos1}
\bar{n}(0,\ep)=\re{\cos\Theta_{n}(0)}.
\e

Far away from the NS interface the Green functions 
aquire their bulk values
\be
\cos\Theta_n(-\infty)=1,\qquad
\cos\Theta_s(\infty)=\frac{\omega}{\sqrt{\Delta^2+\omega^2}}.
\e

%%%%%%%%%%%%%%%%%%%%%%%%%%%%%%%
\begin{figure}
\includegraphics[width=7.5cm]{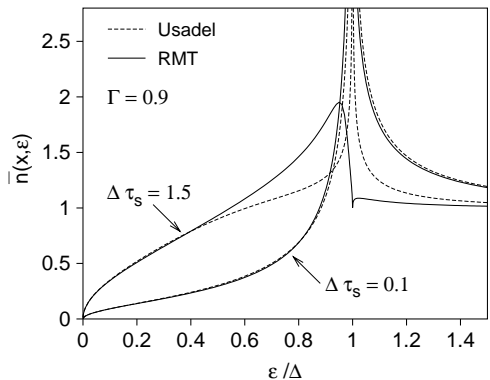}
\caption{
The mean LDOS from the random matrix theory (solid lines),
Eqs.\ (\ref{exciting},\ref{InftymodeLDOS}),
is compared to that from the Usadel 
theory (dashed lines), Eqs.\ (\ref{Uldos1}, \ref{Uresult}),
for the corresponding value of the parameter 
$\gamma_B^2=\Delta\tau_s(2-\Gamma^2)/(2\Gamma)^2$.  
The curves always coincide for small energies 
$\ep\ll \Delta$. The perfect agreement in the 
entire energy range is found in the dirty limit 
$\Delta\tau_s\to 0$, where the Usadel equation is justified.
}
\label{fig:Usadel}
\end{figure}
%%%%%%%%%%%%%%%%%%%%%%%%%%%%%%%%%

The finite transparency of the
interface comes into play in
the appropriate matching conditions at $x=0$\cite{KL} 
\beml
\label{match}
\beq
\label{gBmatch}
\gamma_B\xi_n \Theta'_n(0)&=&\sin\lt[
\Theta_s(0)-\Theta_n(0)\rt],\\
\label{gmatch}
\gamma \xi_n \Theta'_n(0)&=&\xi_s \Theta'_s(0).
\eq
\eml

Once the superconductor is sufficiently clean 
the first term in Eq.\ (\ref{UsadelS}) can be disregarded,
hence $\Theta_s(x)=\Theta_s(\infty)$ and $\Delta(x)=\Delta$
for $x>0$. This justifies the ``rigid'' boundary conditions, which are
used throughout the article.  

The first integral of Eq.\ (\ref{UsadelN}) is readily found
\be
\frac{D_n}{4}(\Theta'_n(x))^2+\omega \cos\Theta_n(x)=\mbox{const},
\e
where the constant is determined from the condition at $x=-\infty$
and equals $\omega$. With the help of Eq.\ (\ref{gBmatch}) 
one obtains\cite{footnote} 
\be
\label{Uresult}
\frac{\sin^2\lt(\Theta_n(0)-\Theta_s(0)\rt)}
{4\gamma_B^2}+\frac{\omega}{\Delta}\lt(
\cos{\Theta_n(0)}-1\rt)=0,
\e
where $\Theta_s(0)$ is substituted
by $\Theta_s(\infty)$ due to the ``rigid''
boundary condition. In the limit $\omega\ll \Delta$
the equation is simplified to
\be
\label{UldosLimit}
\lt(\cos{\Theta_n(0)}-\frac{\omega}{\Delta}\rt)^2=
\frac{4\gamma_B^2\omega}{\Delta}\lt(
1-\cos{\Theta_n(0)}\rt).
\e
Its solution gives rise to the LDOS for $\ep \ll \Delta$
\be
\label{Uscaling}
\bar{n}(0,\ep)=\re\sqrt{-i\frac{4\ep\gamma_B^2}{\Delta}}
\sqrt{1-i\frac{\ep \gamma_B^2}{\Delta}},
\e
which is manifestly equivalent to Eq.\ (\ref{scaling})
and establishes the following relation between the parameters
\be
\label{U2RMT}
\gamma_B^2=\Delta\tau_s \frac{(2-\Gamma)^2}{4\Gamma^2}.
\e 
This relation also follows directly from the definition of $\gamma_B$,
up to a numerical factor, since one can effectively substitute
$R_B=(h^2/e)(2-\Gamma)/2\Gamma$,
$\rho_n=(h^2/e)\ell^{-1}$, and $D_n=\ell^2/\tau_s$.

We conclude that the LDOS obtained from the Usadel equation 
always coincides with that found from Eq.\ (\ref{InftymodeLDOS}) 
for small energies $\ep \ll \Delta$. 
We also demonstrate numerically in Figs. \ref{fig:figTa}, 
\ref{fig:figScI}, and \ref{fig:Usadel} 
that our result for $N\gg 1$ is perfectly consistent 
with the Usadel theory in the dirty limit $\Delta\tau_s\ll 1$,
where the latter is justified. 

One should note, however, that the agreement with 
the quasiclassical theory becomes better with
the increasing barrier height. Indeed, in the 
perfectly transparent interface $\Gamma=1$, the agreement
is reached only in the extremely dirty limit $\Delta\tau_s\to 0$,
while for smaller values of $\Gamma$ the dirty-limit condition 
is less restrictive (see Fig.~\ref{fig:figTa}a).

\section{Conclusion} \label{sc:Conclusion}

In conclusion, we compute the mean LDOS 
in a normal-metal disordered wire in the immediate vicinity 
of an NS interface at zero temperature and zero magnetic field. 
Our calculation is based on the scattering approach and 
takes into account the spatial phase coherence
in the normal-metal. 

We derive the general formula (\ref{gen_result}), 
which expresses the one-point
Green function in terms of the reflection matrices.
The formula can be applied in order to calculate the 
LDOS (and its distribution)
in the wire at arbitrary distance to the NS interface.
In this paper we only consider the mean LDOS at the ballistic 
distance to the interface so that it does not acquire a 
spatial dependence.  

We obtain the relation (\ref{first})
between the disorder-averaged LDOS near the ideal NS interface
and the probability density of the eigenphases of the 
matrix correlator $r_0(\ep)r_0(-\ep)^\dagger$, where
$r_0(\ep)$ is the reflection matrix for the semi-infinite 
normal-metal wire. 

We also study in detail the case of the normal-superconductor
tunnel junction and derive the self-consistent equation 
(\ref{InftymodeLDOS}) that determines the LDOS 
in the diffusive normal metal. In the 
dirty limit our expression coincides with the LDOS found by Golubov and
Kupriyanov\cite{GK96} from the Usadel equation. 

The quasiclassical analysis
of the Green function at the NS interface of finite
transparency has been performed by many authors
\cite{KL,GK89,GK96,Z1,Z2} in connection with the
boundary conditions of semiclassical superconductivity.
However, to our best knowledge 
there exists no counterpart to the equation (\ref{InftymodeLDOS})
in the literature. 

In the case of an ideal NS interface the LDOS is found to be
almost independent of the number of channels in a wire,
except for very small energies, hence its 
insensitivity to phase-coherence effects. This persists 
to the case of finite transparency provided
the clean limit condition $\Delta\tau_s \gg 1$.
In the dirty limit $\Delta\tau_s \ll 1$ and small
transparency $\Gamma \ll 1/N$ the situation is different
and the phase-coherent effects play a role.

The effect of Anderson localisation is seen 
in the linear increase of the LDOS, $\bar{n}=
(\pi/4)(N+1)\ep\tau_s(2-\Gamma)/\Gamma$,
for energies lower than $\ep_c=1/N^2\tau_s$.
In the diffusive metal, $N\to \infty$,  
the LDOS increases as the square root of energy 
$\bar{n}=\re \sqrt{-i\ep \tau_s}(2-\Gamma)/\Gamma$. 
The form of the crossover in energy dependence 
of the LDOS from linear to square-root 
behavior is given by Eq.\ (\ref{result0}) for weak disorder 
and perfect NS interface.

\begin{acknowledgments}
We thank C.~W.~J.~Beenakker, P.~W.~Brouwer and R.~Narayanan
for helpful discussions. We are especially grateful
to Alexander Golubov for bringing the results of 
Ref.\ \onlinecite{GK96} 
to our attention.
\end{acknowledgments}

\end{document}